\documentclass{optica-article}
\usepackage{color}

\journal{opticajournal} % for journals or Optica Open

\articletype{Research Article}

\usepackage{lineno}
\begin{document}

\title{%Spectroscopy and modal structure of Lithium Tantalate whispering gallery resonators
Accurate measurement of Lithium Tantalate birefringence in whispering gallery resonators
%\textit{alternative title:}\\
%Resolving the Sign of Birefringence in Lithium Tantalate\\
%\textit{other suggestions welcome!}
}

\author{Linjie Shao,\authormark{1,2} Andrey B. Matsko,\authormark{3} Anatoliy A. Savchenkov,\authormark{3}  Harald G. L. Schwefel,\authormark{1,2} Nicholas J. Lambert,\authormark{1,2} Florian Sedlmeir,\authormark{1,2} Dmitry V Strekalov\authormark{1*}}

\address{
\authormark{1}Department of Physics, University of Otago, 730 Cumberland Street, Dunedin 9016, New Zealand\\
\authormark{2}The Dodd-Walls Centre for Photonic and Quantum Technologies, Dunedin, New Zealand\\
\authormark{3}Jet Propulsion Laboratory, California Institute of Technology, 4800 Oak Grove drive, Pasadena, USA}

\email{\authormark{*}strdm578@otago.ac.nz} %% email address is required; 

% use {asbstract*} to suppress the copyright line. Copyright information will be added in production

\begin{abstract*} 
Lithium Tantalate (LTA) is an important ferroelectric crystal widely used in piezoelectric, electro-optic, and nonlinear optical devices. Although this crystal is very common, there is an ongoing debate regarding its birefringence magnitude, and even its sign. While some sources claim it to be negative, others maintain that it is positive. The issue is further complicated by the sample-to-sample variation in the crystal composition, ranging from congruent to stoichiometric. Here we report the experimental exploration of this question based on high-resolution spectroscopy of optical whispering gallery resonators fabricated from this crystal. These sensitive measurements unambiguously indicate that in the near-infrared range at room temperature the stoichiometric LTA is negative, while the congruent LTA is positive.
\end{abstract*}

%%%%%%%%%%%%%%%%%%%%%%%%%%  body  %%%%%%%%%%%%%%%%%%%%%%%%%%
\section{Introduction}

LTA is widely used in acousto- and electro-opical modulators, waveguides, piezoelectric and nonlinear-optics devices, as well as in micro-electro-mechanical systems (MEMS). In spite of this wide range of applications, birefringent properties of this crystal are not sufficiently well explored, and the reports on the magnitude and even the sign of its birefringence are contradictory. For example, a study of the temperature-dependent Sellmeier equations \cite{Abedin1996LiTaO3} indicates that this crystal is positive at room temperature, i.e. $n_e>n_o$, where $n_e$ and $n_o$ stand for the extraordinary and ordinary refractive indexes, respectively. However a different experimental study \cite{nakamura2002SLT} showed that while the congruent LTA is indeed positive in the visible optical range, the stoichiometric LTA is negative, i.e. $n_e<n_o$. The distinction arises because in LTA $n_e$ is affected by nonstoichiometric defects (as later confirmed by  \cite{kostritskii2009LTA}), and $n_o$ is not. The work by \cite{nakamura2002SLT} also shows that magnesium-oxide doping up to 1 mol\% has no detectable effect on the stoichiometric LTA refractivity in this range. A later study of magnesium-oxide doped stoichiometric LTA \cite{Dolev2009LTa} confirmed that this crystal has negative birefringence, yielding a different pair of Sellmeier equations. 

A crystallographic study \cite{huband2017LiNbTaO3,huband2017LTA} of LTA and composite Lithium niobate-tantalate crystals does not provide Sellmeier equations, but leads to an interesting insight into the temperature dependence of these materials birefringence. It shows that at low temperatures the crystals are generally negative, and at high temperatures, positive. There exist a temperature in-between when the birefringence completely vanishes and the crystal becomes optically isotropic. For a pure congruent LTA this zero-birefringence temperature is estimated to be slightly above the room temperature for the visible optical range, hence this crystal is expected to be weakly positive at room temperature.  

In this research we employ the optical spectroscopy of high-$Q$ whispering gallery mode (WGM) resonators to investigate the birefringence of two stoichiometric and one congruent LTA samples. The motivation for this research is two-fold. First, its material-science aspect is intriguing in itself, potentially finalizing the debate regarding LTA having positive or negative birefringence. Second, this knowledge has a practical importance in the context of LTA WGMs application such as the recently proposed electro-optical converter \cite{strekalov2022Receiver}. LTA appears to be a particularly promising crystal for this and similar applications, significantly surpassing lithium niobate in terms of the resonator optical quality factor $Q$ and therefore predicting a higher conversion efficiency. The modelling of the underlying phase matching for the electro-optical converter requires a detailed and accurate knowledge of the refraction properties of this crystal, as will be described below.

WGM resonator optics not only makes use of high-quality optical materials, but also provides a very sensitive technique for their study. Previously, we used this technique to observe the photorefractivity of lithium niobate in weak infrared light, particularly in MgO-doped crystals \cite{savchenkov06photorefraction,savchenkov06MgOLiNbO3,savchenkov07damage}. Here we use it to explore the modal structure and birefringent properties of crystalline WGM micro-resonators. Specifically, we employ the sideband spectroscopy technique discussed in \cite{Li12sb,Savchenkov12overmoded,Strekalov14SFG}. This technique allows for a very precise (better than one part in 10,000, in our case) determination of the free spectral range (FSR) of a targeted mode family. As a result, the modes can be identified by their radial order $q=1,2,3,...$ important for the phase matching in a variety of nonlinear optics processes. Knowing $q$ we can also find a theoretical FSR value as explained below. Using refractive index as a free parameter and comparing the theoretically predicted value to the measured one we can infer the refractive properties of the crystal from which the resonator is made.

\section{Fabrication and coupling of resonators}

Two resonators were fabricated from a congruent and a stoichiometric LTA samples at the University of Otago, and one, also stoichiometric, at JPL. The stoichiometric samples were doped with 1 mol\% of magnesium oxide. They both originated from the Oxide Corporation but came from different batches.

\begin{figure}[htbp]
\centering%\includegraphics[angle=-90,origin=c,width=3cm]{IMG_0571}
\includegraphics[width=10cm]{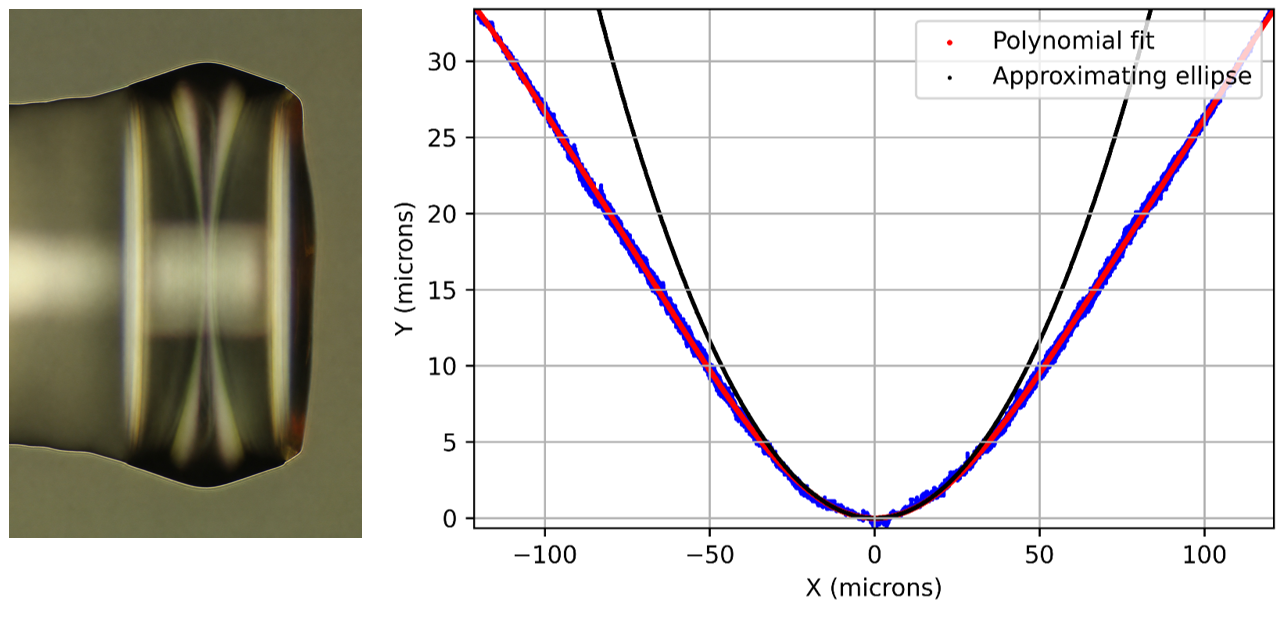}
\caption{A photo of the JPL LTA resonator with $R=429\,\mu$m (left), and its rim profile plot fitted to a polynomial and ellipse (right). The aspect ratio is not preserved in the plot.}
\label{fig:W-5}
\end{figure}

The resonator fabrication process included single-point diamond turning followed with diamond slurry polishing, a process described in e.g. \cite{grudinin06highQlin,grudinin06highQ,fujii2020fab,minkov2021Fabrication}. A photo of the JPL resonator is shown in Fig. \ref{fig:W-5}. This photo allows us to determine the resonator diameter with a single-micron precision and to find its radius $R=429\,\mu$m. Determining the rim radius is slightly more complicated. It involves converting the image of the rim boundary to a 2D data set, which is then fitted with a polynomial and an ellipse, as shown in Fig. \ref{fig:W-5}. The elliptical fitting is necessary because the theoretical model \cite{Demchenko13} we use for the WGM dispersion is developed for ellipsoidal resonators. From Fig. \ref{fig:W-5} we see that this approximations very closely follows the actual rim outline in the equatorial region where the optical modes are located. The fit yields the rim radius $r=109\,\mu$m.

The University of Otago resonators were shaped by diamond cutting or femtosecond-laser machining, followed by mechanical polishing of the resonator rim. The resulting resonators had significantly larger radii than the JPL resonator -- $R \approx 2.3~\mathrm{mm}$ for the stoichiometric sample and $R \approx 1.6~\mathrm{mm}$ for the congruent sample. In all measurements, the resonators were thermally stabilized and coupled to the respective pump lasers using the standard frustrated total internal reflection approach, see e.g. \cite{Gorodetsky99coupling,foreman2016dielectric,sedlmeir_polarization-selective_2017,trainor2018selective}. The pump wavelength was centered at 1558 nm for the JPL measurement, and at 1550 nm for the University of Otago measurement. The laser light travels down a single-mode fiber, where its polarization can be arbitrary set with a fiber polarization controller. Upon exiting the fiber, light is collimated by a graded index (GRIN) lens and is focused at the spot where the resonator surface approaches the diamond prism coupler, enabling the near-field coupling. A similar pigtailed GRIN lens collects the reflected light into a single-mode fiber coupled to a photodiode.  

\begin{figure}[htbp]
\centering\includegraphics[width=13cm]{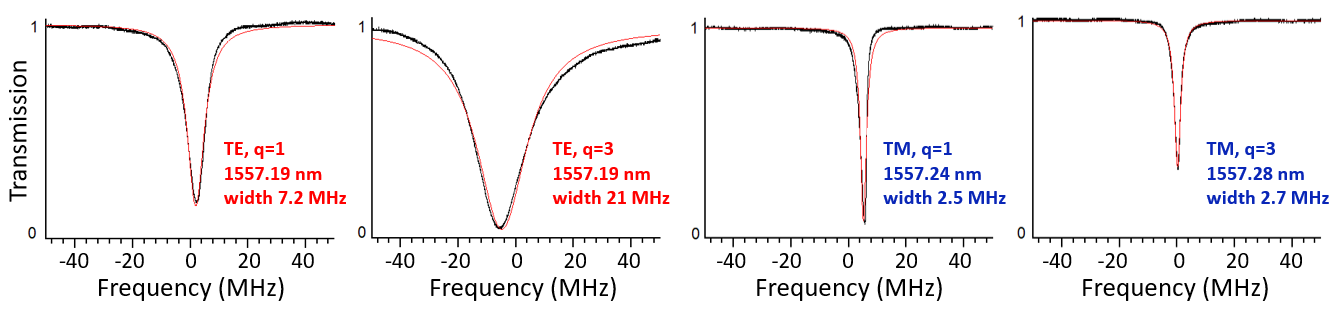}
\caption{Examples of WGM resonance frequency spectra for TE and TM polarizations and radial mode numbers $q=1$ and $q=3$ in the $R=429\,\mu$m resonator. The red line is a Lorentzian fitting curve.}
\label{fig:lines}
\end{figure}

The resonator modes are visible in the input-output power transfer function as narrow dips. When the free-space optics are optimized to match the WGM emission angle and spatial profile, and if the critical coupling is achieved, the contrast of these dips approaches unity. In Fig. \ref{fig:lines} we show the examples of such modes observed in the JPL resonator. Two of the TE (extraordinary polarized, E-field is parallel to the resonator symmetry axis) and two of the TM (ordinary polarized) families, with the radial numbers $q=1$ and $q=3$. Lorentz-fitting provides the resonance bandwidths and the quality factors $Q$. We see that the TM modes have significantly higher $Q$ than the TE modes. We also see that the incidence angle inside the coupler is optimized for higher-order TE modes and fundamental ($q=1$) TM modes. Indeed, the TE $q=3$ mode is clearly over-coupled compared to the $q=1$ mode, whereas the TM $q=3$ mode is clearly under-coupled compared to the $q=1$ mode, which appears to be near the critical coupling.
 
\begin{figure}[htbp]
\centering\includegraphics[width=13cm]{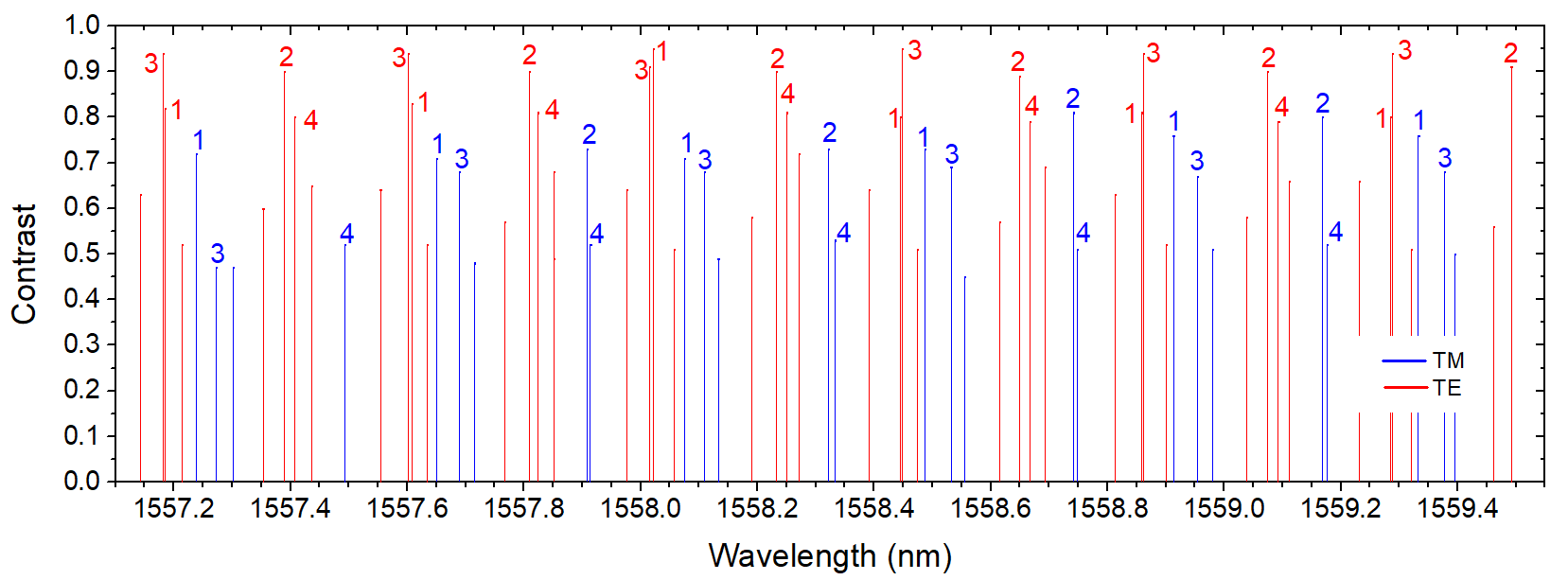}
\caption{% From FSR and Spectrum22C.opj
Schematic representation of the TE and TM WGM spectrum in the JPL resonator, reflecting the modes coupling efficiency and wavelengths. Numbers represent the radial order $q$ determined from the frequency-modulation spectroscopy.}
\label{fig:Spectrum}
\end{figure}

\begin{figure}[htbp]
\centering\includegraphics[width=12cm]{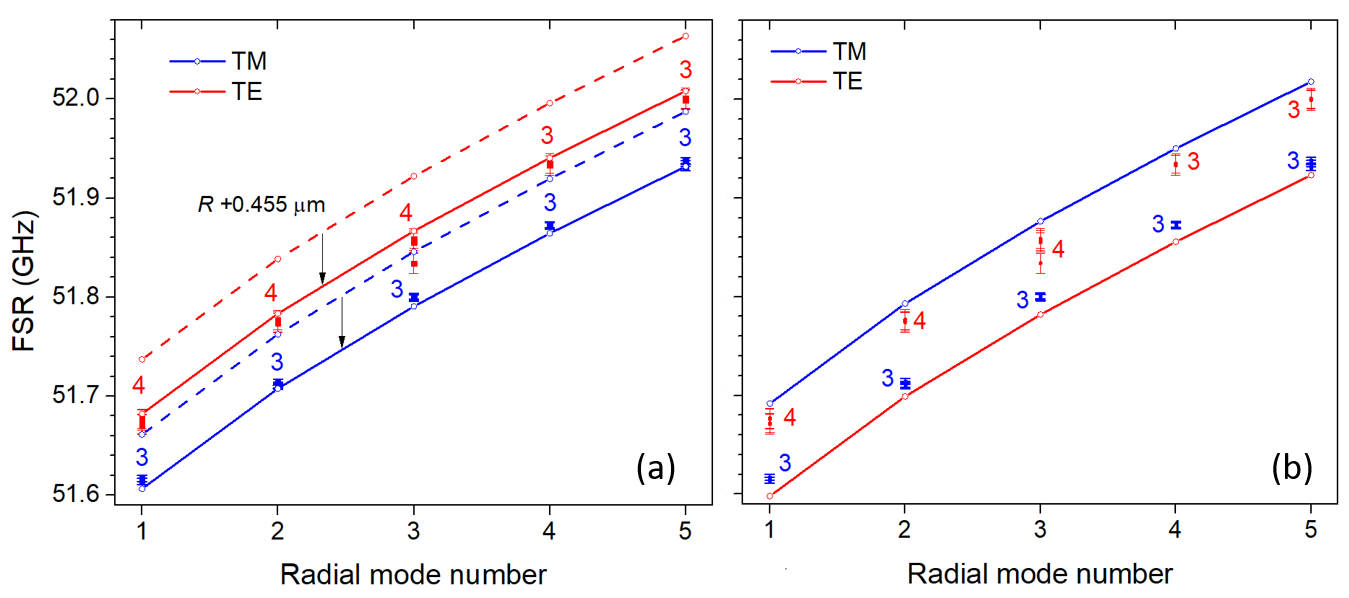}
\caption{% From 02-28-24/FSR.opj
Dots with error bars are the measured FSR values for various WGM families in the JPL resonator from Fig. \ref{fig:W-5}. The numbers indicate how many data points are in each small cluster. Dashed and solid lines in plot (a) represent the theoretical results before and after the radius correction, respectively. They are based on the negative-birefringence Sellmeier equations \cite{Dolev2009LTa}. The lines in plot (b) represent the theoretical results based on the positive-birefringence Sellmeier equations \cite{Abedin1996LiTaO3}.
}
\label{fig:TETMFSR}
\end{figure}

The WGM resonator spectrum observed in this way consists of a large number of TE and TM modes, as shown in Fig. \ref{fig:Spectrum}. The spectral analysis readily provides their coupling efficiency and relative position on the optical wavelength scale, but does not tell us the radial order of each mode. Knowing the radial order is important for nonlinear optics applications. In order to identify it we measure FSRs of a large number of modes shown in Fig. \ref{fig:Spectrum} using the sideband spectroscopy technique \cite{Li12sb,Savchenkov12overmoded,Strekalov14SFG} and build a histogram of their values. We find that the FSR values for each polarization family form very tight clusters, such as can be seen in Fig. \ref{fig:TETMFSR}. These clusters correspond to the radial mode numbers, with the minimum FSR indicating the $q=1$ modes, and so on. This is how the modes in Fig. \ref{fig:Spectrum} were identified and labelled. Note that the azimuthal mode order, $p=L-m$, does not affect the FSR within our measurement accuracy. This is why we find multiple elements for the same $q$ cluster even when probing modes within a single FSR range.

It should be pointed out that in large resonators with high spectral density of modes their cross-coupling leads to spectral irregularities that may affect the FSR measurements, as has been reported in, e.g., \cite{Strekalov14SFG}. Such WGMs stand out due to their distorted line shapes and reduced coupling contrast. They have been avoided in our measurements as purely resonator phenomena irrelevant to the material dispersion.

Shown in Fig.~\ref{fig:TETMFSR}(a) and (b) along with the same experimental data set are the theoretical results, whose derivation will be explained in the following. The calculations for Fig.~\ref{fig:TETMFSR}(a) are based on the Sellmeier equations derived by \cite{Dolev2009LTa} and recommended by the Oxide Corporation for their LTA. According to these equations, LTA is a \textit{negative} crystal with $n_e<n_o$. The initial estimates shown as dashed lines lead to a discrepancy between the theoretical and experimental results, uniform for both TE and TM polarizations and all radial mode numbers. This can be due to the the radius measurement error. Compensating this error by only a $0.455\,\mu$m radius correction leads to the adjusted theoretical estimate, shown as solid lines. Note that this radius correction $R \rightarrow R+0.455\,\mu$m is well within our optical measurement accuracy based on analysis of the image from Fig. \ref{fig:W-5}, and is therefore plausible.

Calculations performed for Fig. \ref{fig:TETMFSR}(b) are based on the Sellmeier equations by \cite{Abedin1996LiTaO3}, according to which LTA is a \textit{positive} crystal with $n_e>n_o$. Here we again adjusted the resonator radius for the best data fit, in this case by $R \rightarrow R-1.2\,\mu$m. However now the discrepancy is persistent and cannot be reconciled by correcting the resonator radius. We conclude that the Sellmeier equations by \cite{Dolev2009LTa} describes the JPL LTA sample fundamentally better than the ones by \cite{Abedin1996LiTaO3}, and that this sample is in fact a negative crystal.

\begin{figure}[htbp]
\centering\includegraphics[width=7.5cm]{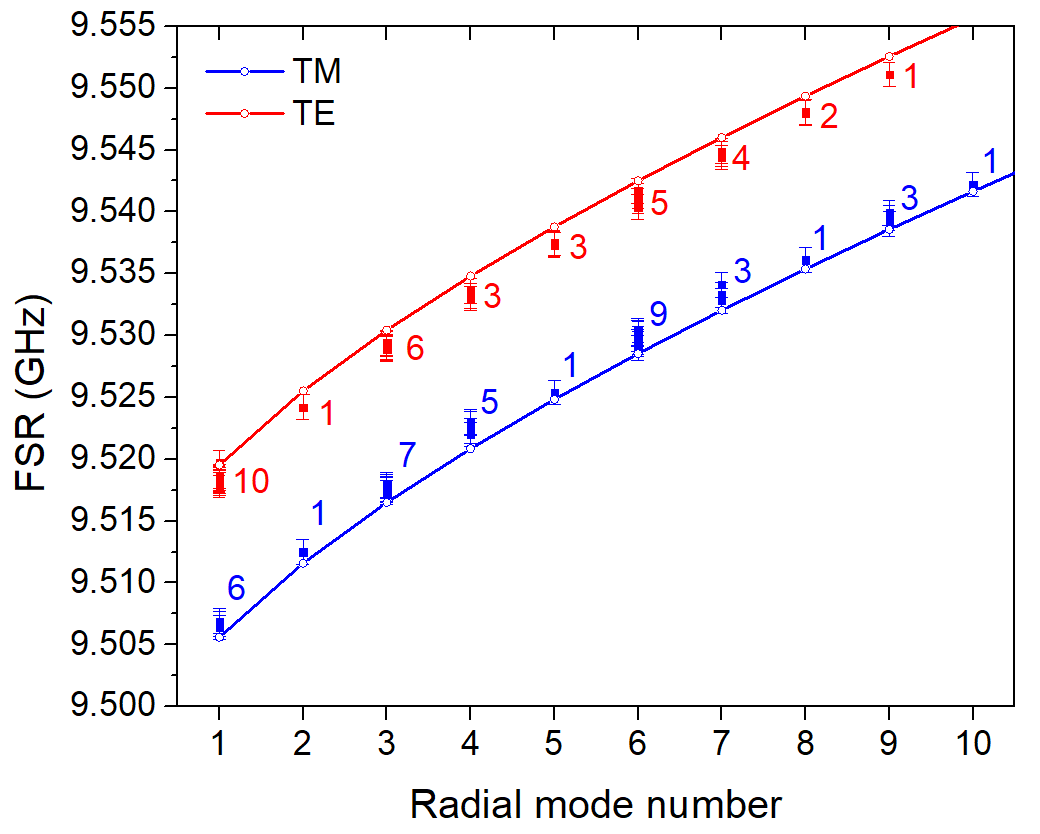}
\caption{% From 02-28-24/FSR.opj
Fitting the measured TE and TM FSRs to the theoretical model derived from the negative-birefringence Sellmeier equations \cite{Dolev2009LTa}. Indicated are numbers of detected modes in each $q$ cluster.
}
\label{fig:TETMFSR1}
\end{figure}

A similar conclusion can be drawn for the Otago resonator. In Fig. \ref{fig:TETMFSR1} we plot the experimental and theoretical results for the TM and TE FSRs in this resonator. The theoretical results are again based on the negative-birefringence Sellmeier equations \cite{Dolev2009LTa}, and the resonator radius is corrected to the best-fit value $R=2.3273\pm 0.0001$ mm. The error margin if calculated based on the number of the $q=1$ data points and the individual FSR measurement accuracy. 

The TE and TM fit can be made individually better if we allow the radius to take different values $R_{\rm TM}=2.3270$ mm and $R_{\rm TE}=2.3276$ mm for the TM and TE $q=1$ clusters, respectively. The small difference $\Delta R=R_{\rm TM}-R_{\rm TE}\approx-0.6\,\mu$m can be attributed to the additional birefringence $\delta n$ unaccounted for by the adopted Sellmeier equations. Representing the total birefringence $\Delta n=n_o-n_e=\Delta n_S+\delta n$, where $\Delta n_S=1.9\times 10^{-3}$ is found from the Sellmeier equations \cite{Dolev2009LTa} and built into out model, we can use a linear proportion
\begin{equation}\label{dndR}
    \frac{2\delta n}{n_o+n_e}=\frac{\Delta R}{R}
\end{equation}
to find the unaccounted part $\delta n$. This results into $\delta n =(-5.5\pm 0.9)\times 10^{-4}$ and leads to $\Delta n=(1.35\pm 0.09)\times 10^{-3}$. 

\begin{figure}[htbp]
\centering\includegraphics[width=7.5cm]{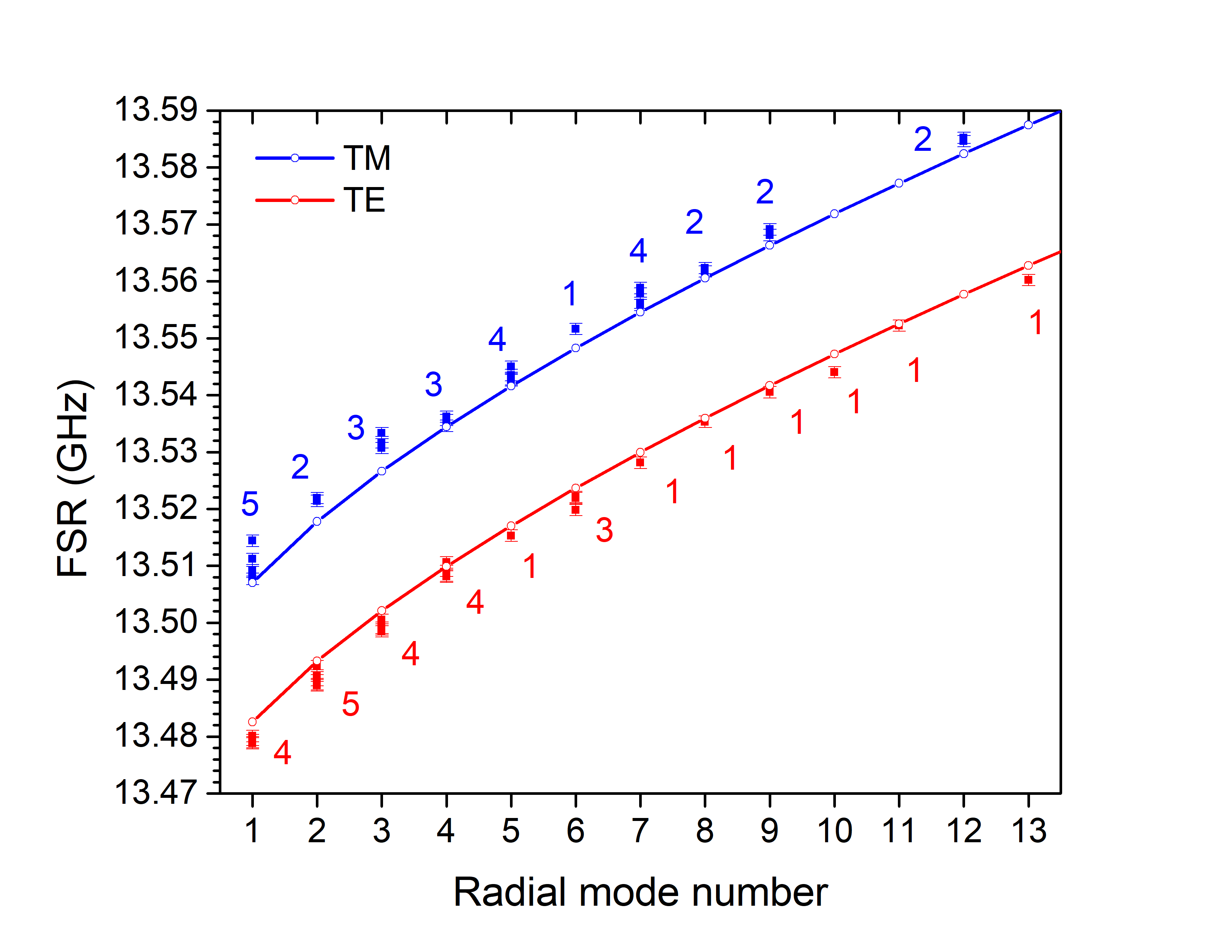}
\caption{
The TM and TE FSR measurements in a congruent LTA resonator show the positive birefringence. Solid lines represent theoretical fit for $R=1.63457$ mm. 
}
\label{fig:congr}
\end{figure}

The measurement performed with the congruent LTA crystal yields a fundamentally different result, shown in Fig. \ref{fig:congr}. In this case the TM FSR is consistently larger than the TE FSR, which indicates positive birefringence. The theoretical fitting in this case is based on the Sellmeier equations derived by \cite{Abedin1996LiTaO3}. The improved resonator radius estimate calculated as before gives $R=1.63457\pm 0.00008$ mm, compromising between the best-fit values $R_{\rm TM}=1.63418$ mm and $R_{\rm TE}=1.63497$ mm for the TM and TE $q=1$ clusters, respectively. Repeating the arguments that have lead to equation (\ref{dndR}) we can estimate the corrected birefringence of this material $\Delta n=\Delta n_S+\delta n$. Now $\Delta n_S=-4.23\times 10^{-3}$ from the Sellmeier equations \cite{Abedin1996LiTaO3} and $\delta n=(-1.02\pm 0.1)\times 10^{-3}$, which leads to $\Delta n=(-5.2\pm 0.1)\times 10^{-3}$. In the following section we will discuss a more rigorous approach to estimating the unaccounted birefringence using the JPL resonator as an example.

\section{Theory}

Our theoretical calculations of the FSR are based on the WGM dispersion model 
\begin{equation}\label{EBK_quartic1}
    k_{L,q,p}\,a=L+\alpha_q\left(\frac{L}{2}\right)^{1/3}+\frac{2p(a-b)+a}{2b}-\frac{n^{\xi}}{\sqrt{n^2-1}}+O(L^{-1/3})
\end{equation}
developed by \cite{Demchenko13}. Here $k_{L,q,p}=\omega_{L,q,p}n(\omega_{L,q,p})/c$ is the mode's eigenvalue, $a=R$ and $b$ are the large and small semiaxes of the ellipsoid approximating the resonator, $\alpha_q>0$ is the Airy function $q^{\rm th}$ zero, and $L\gg 1$ is the mode's orbital number. Parameter $\xi$ in the last term of the right-hand side of this equation takes on value +1 for the extraordinary polarized TE mode and -1 for the ordinary polarized TM mode. The corresponding extraordinary or ordinary refraction indices $n_{\rm e,o}$ depend on the mode's eigenfrequency, which makes this equation transcendent. But it can be solved for an eigenfrequency $\omega_{L,q,p}$ numerically for any orbital momentum $L$ relying on the $n_{\rm e,o}(\omega)$ dependence expressed by the Sellmeier equations. Then the FSR is found as $\Omega=\omega_{L+1,q,p}-\omega_{L,q,p}$. These are the theoretical values shown in Fig. \ref{fig:TETMFSR}(a) and (b) based on different Sellmeier equations.

A fully conclusive verification of the WGM dispersion equation (\ref{EBK_quartic1}) is difficult because of the measurement uncertainties of the parameters $a$ and $b$ and sample-to-sample variation of the Sellmeier equation parameters. However multiple partial tests indicate that this equation can be trusted. For example, this is attested to by a successful modelling of the mutual arrangement of various spectral lines, enabling their mode-number identification \cite{Schunk14modes}. In the present work, we observed the lack of the FSR dependence on the orbital $p$-number, which is also consistent with equation (\ref{EBK_quartic1}). It is also easy to see from Fig. \ref{fig:TETMFSR} that the TM to TE FSR ratio remains approximately constant for all radial mode numbers $q$. This observation is also consistent with the dispersion equation. It will allow us to make a fine correction to the material birefringence predicted by the Sellmeier questions.

Expanding $n_{\rm e,o}(\omega+\Omega)-n_{\rm e,o}(\omega)\approx \Omega n_{\rm e,o}^\prime(\omega)$ and discarding other small terms, we arrive at
\begin{eqnarray}\label{TMTEFSR}
    \Omega_{\rm TM}\frac{R}{c}\left[n_o+\omega n^\prime_o-K_o\right]&\approx&1+\frac{\alpha_q}{3}\left(2L^2\right)^{-1/3},\\
    \Omega_{\rm TE}\frac{R}{c}\left[n_e+\omega n^\prime_e-K_e\right]&\approx&1+\frac{\alpha_q}{3}\left(2L^2\right)^{-1/3},\nonumber
\end{eqnarray}
where 
\begin{equation}\label{Koe}
    K_o=\frac{c}{R}\frac{n^\prime_o}{n_o^2}\frac{n_o^2+n_o-1}{(n_o^2-1)^{3/2}}\quad {\rm and}\quad
K_e=\frac{c}{R}n^\prime_e\frac{n_e^2-n_e+1}{(n_e^2-1)^{3/2}}.
\end{equation}
Dividing equations (\ref{TMTEFSR}) for the same $q$ we arrive at
\begin{equation}\label{FSRratio}
    \frac{\Omega_{\rm TM}}{\Omega_{\rm TE}}=\frac{n_e}{n_o}\frac{1+\omega \frac{n^\prime_e}{n_e}-\frac{K_e}{n_e}}{1+\omega \frac{n^\prime_o}{n_o}-\frac{K_o}{n_o}}\approx 0.99971\frac{n_e}{n_o}.
\end{equation}
Here the numeric factor is evaluated for the JPL resonator based on the Sellmeier equations from \cite{nakamura2002SLT}. It captures the effects of the group velocity dispersion (GVD) and the resonator's evanescent field. Evidently, these effects are small, so this number can be treated as a constant insensitive to the small birefringence correction we are about to derive. 

\begin{figure}[htbp]
\centering\includegraphics[width=7cm]{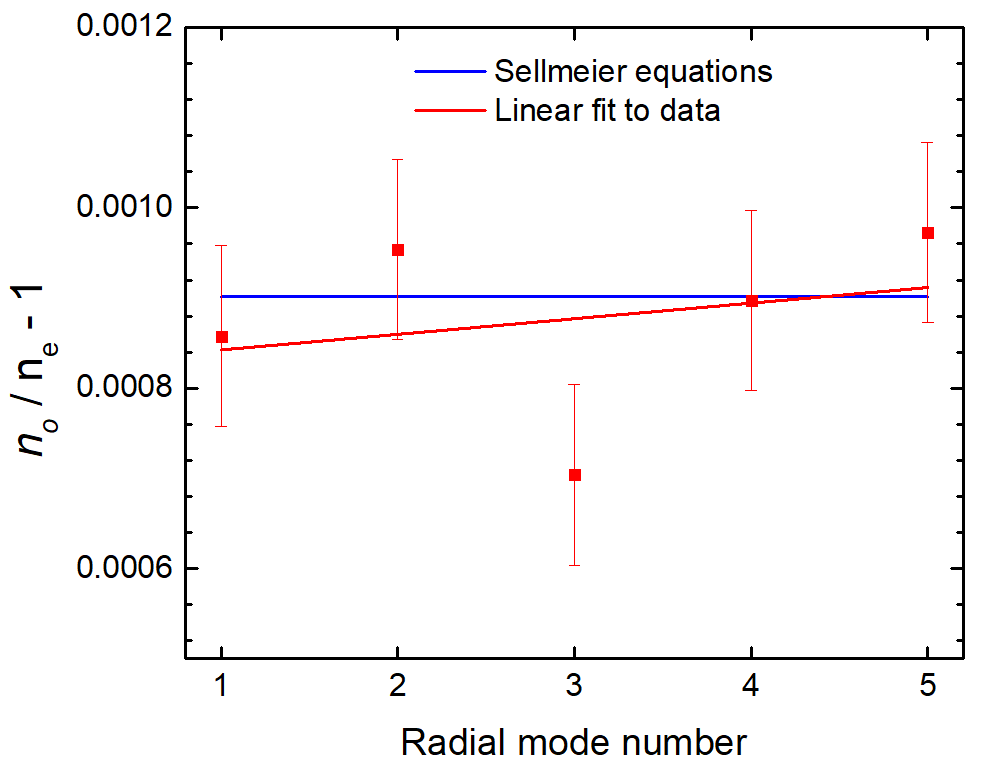}
\caption{The birefringence calculated from the Sellmeier equations \cite{Dolev2009LTa} and based on the FSR ratio measurement in the JPL resonator.}
\label{fig:FSRratio}
\end{figure}

A weak birefringence can be conveniently parametrized as e.g. $\Delta n/n_e= n_o/n_e-1$. In Fig. \ref{fig:FSRratio} we plot this value derived from the Sellmeier equations \cite{Dolev2009LTa} alongside with the experimental result based on the measured FSRs and equation (\ref{FSRratio}). From the experimental data we see that indeed, the FSR ratio is independent of the radial mode number within the error margin. We also see that it reliably yields a negative birefringence closely matching the Sellmeier equations prediction: $\Delta n =(1.8\pm 0.1)\times 10^{-3}$. Similar analysis can be performed for the other two resonators.

Our analysis so far was based on the assumption that the TE-polarized modes ``see'' exclusively the extraordinary index of refraction, and the TM modes, the ordinary. This is a very good approximation for large resonators. However at our level of sensitivity we may start seeing the effect of the minor electric field projections, such as the extraordinary ($\hat{e}_\theta$) projection for TM and ordinary ($\hat{e}_r$ and $\hat{e}_\phi$ ) projections of TE modes. Let us estimate how significantly this affects our FSR measurements. 

It is reasonable to evaluate the effective FSR as the weighted average of those found for purely ordinary and extraordinary polarizations
\begin{equation}\label{wt_avg}
\Omega_{\rm eff}\approx\frac{\Omega_{\rm TE}W_{\rm e}+\Omega_{\rm TM}W_{\rm o}}{W_{\rm e}+W_{\rm o}},
\end{equation}
where the weights $W_{\rm e}$ and  $W_{\rm o}$ are proportional to the electric field energy associated with the electric field energy for the corresponding polarization:
\begin{equation}\label{energy}
W=\frac{n^2}{8\pi}\int \left|\vec E\right|^2\,\sin\theta r^2 dr d\theta d\phi.
\end{equation}
In spherical resonators, the fields can be found from the Debye potential $\psi$ as \cite{Oraevsky02WGM} 
\begin{eqnarray}\label{DebyeSp}
\vec{E}_{TE}&=&\frac{\hat{e}_\theta}{r\sin\theta}\frac{\partial\psi}{\partial \phi}-\frac{\hat{e}_\phi}{r}\frac{\partial\psi}{\partial \theta}\\
\vec{E}_{TM}&=&\frac{i}{k_0}\left(\hat{e}_r\frac{L(L+1)}{n_o^2r^2}\psi+\frac{\hat{e}_\theta}{n_e^2r}\frac{\partial^2\psi}{\partial r\partial \theta}+\frac{\hat{e}_\phi}{n_o^2r\sin\theta}\frac{\partial^2\psi}{\partial r\partial \phi}\right),\nonumber
\end{eqnarray}
Here we modified the result by \cite{Oraevsky02WGM} to make sure that the ordinary field projections are associated with the ordinary index of refraction $n_o$, and the extraordinary, with $n_e$. We also need to keep this in mind when evaluating the field energy (\ref{energy}). As a good approximation, we can write 
\begin{equation}
\psi(\vec r)=\psi_0\psi_r(r)\psi_\theta(\theta)\frac{e^{im\phi}}{\sqrt{2\pi}}
\end{equation}
where the integral of each function's square over the respective coordinate is normalized to unity, and $\psi_0$ is a constant. Note that $\partial/\partial\phi=im$.

Let us consider the TE case. From the first equation of (\ref{DebyeSp}) and (\ref{energy}) it is easy to find 
\begin{equation}
W_{\rm e}\approx\psi_0^2\frac{n_e^2m^2}{8\pi R^2},
\end{equation}
where the approximation is associated with replacing the variables $r$ and $\theta$ outside the Debye potential arguments with constants $R$ and $\pi/2$, respectively. Finding  $W_{\rm o}$ requires the evaluation of $\partial \psi_\theta/\partial\theta$ and integration of its square. We will use the approach by \cite{Breunig13eigenfunc} modeling $\psi_\theta$ as a Gauss-Hermite function
\begin{equation}\label{IngoEqn6}
\psi_{\theta}=\left(\sqrt{\pi}2^pp!\right)^{-1/2}e^{-\eta^2/2}H_p(\eta),\quad {\rm where}\quad \eta=\sqrt{m}\left(R/{\rm r}\right)^{1/4}(\pi/2-\theta).
\end{equation} 
We find this function's derivative and take advantage of the relation
\begin{equation}
\eta H_p(\eta)-H_p^\prime(\eta)=\frac{1}{2}H_{p+1}(\eta)-pH_{p-1}(\eta),
\end{equation}
as well as of the orthogonality relations, to arrive at
\begin{equation}\label{Tintegral}
\int_0^\pi\left(\frac{\partial \psi_\theta(\theta)}{\partial \theta} \right)^2\,d\theta\approx\sqrt{m}\left(\frac{R}{\rm r}\right)^{1/4}\left(p+\frac{1}{2}\right).
\end{equation}
Consequently,
\begin{equation}
W_{\rm o}\approx\psi_0^2\frac{n_o^2\sqrt{m}}{8\pi R^2}\left(\frac{R}{\rm r}\right)^{1/4}\left(p+\frac{1}{2}\right).
\end{equation}
Now we can evaluate the corrected FSR value for the TE modes based on (\ref{wt_avg}):
\begin{equation}
\Omega_{\rm TE\,eff}\approx\Omega_{\rm TE}+\left(\Omega_{\rm TM}-\Omega_{\rm TE}\right)
\left(p+\frac{1}{2}\right)\left(\frac{n_o}{n_e}\right)^2\left(\frac{R}{\rm r}\right)^{1/4}m^{-3/2}.
\end{equation}
We see that because of the factor $m^{-3/2}$ this correction is very small compared to the observed FSR birefringence $\Omega_{\rm TM}-\Omega_{\rm TE}$, and would not be detectable in our experiment. 

Let us now consider the TM case. From the second equation of (\ref{DebyeSp}) and (\ref{energy}) we see that this case requires evaluating and integration of not only $\partial \psi_\theta/\partial\theta$, but also $\partial \psi_r/\partial r$. For this part of the eigenfunction we use the Airy function approximation \cite{Breunig13eigenfunc}
\begin{equation}
\psi_r(r)\approx \frac{(2m)^{1/6}{\rm Ai}\left((2m)^{1/3}w/R-\alpha_q\right)}{R^{3/2} {\rm Ai}^\prime\left(-\alpha_q\right)}, 
\end{equation}
where $-\alpha_q$ is the Airy function's $q$th zero and the coordinate $w\approx R-r$ is measured from the resonator surface inwards. The relation
\begin{equation}
\int_{-\alpha_q}^\infty {\rm Ai}^2(x)\,dx= \left({\rm Ai}^\prime\left(-\alpha_q\right)\right)^2
\end{equation}
was used to calculate the normalization factor. A similar relation
\begin{equation}
\int_{-\alpha_q}^\infty\left(\frac{\partial}{\partial x} {\rm Ai}(x)\right)^2\,dx= \frac{\alpha_q}{3}\left({\rm Ai}^\prime\left(-\alpha_q\right)\right)^2
\end{equation}
allows us to find 
\begin{equation}
\int_0^\infty\left(\frac{\partial \psi_r(r)}{\partial r} \right)^2\,dr= \frac{\alpha_q(2m)^{2/3}}{3R^4}.
\end{equation}
This and the earlier found integral (\ref{Tintegral}) lead to
\begin{equation}
W_{\rm e}\approx\psi_0^2\frac{\alpha_q(p+1/2)}{24\pi k_0^2n_e^2 R^4}\left(\frac{R}{\rm r}\right)^{1/4}2^{2/3}m^{7/6}.
\end{equation}
The expression for $W_{\rm o}$ consists of two parts, one arising from the $\hat{e}_r$ field and the other from $\hat{e}_\phi$ field. The second part is small compared to the first and can be neglected. With this, we arrive at the  corrected FSR value for the TM modes:
\begin{equation}
\Omega_{\rm TM\,eff}\approx\Omega_{\rm TM}+\left(\Omega_{\rm TE}-\Omega_{\rm TM}\right)
\alpha_q\left(p+\frac{1}{2}\right)\left(\frac{n_o}{n_e}\right)^2\left(\frac{R}{\rm r}\right)^{1/4}\frac{2^{2/3}m^{7/6}}{3L^4}.
\end{equation}
Once again, the factor multiplying the observed FSR birefringence $\Omega_{\rm TE}-\Omega_{\rm TM}$ is very small, of the order $O\left(m^{-17/6}\right)$, and cannot be detected in our measurement. We conclude that any discrepancy between the experimental and theoretical FSR difference must be due to the material birefringence uncertainty. This conclusion based on the analytical approach above, is further confirmed by the finite-element model simulations.

\section{Role of the birefringence in electro-optic phase matching}

The results of the current study are important for optimization of all-dielectric photonic converters and receivers \cite{strekalov2022Receiver}. The sensitivity of the photonic receiver is determined not only by the quality factors of the optical resonances and the resonant enhancement of the strength of the optical and microwave fields inside the resonator, but also by the ability of the microwave field to efficiently couple distinct optical mode families.  The receiver can be viewed as an electro-optic frequency converter in which an applied microwave field produces coupling between optical WGMs whose frequency separation is close to the microwave frequency.  In this picture, phase matching is an essential condition for obtaining a strong electro-optic interaction. The importance of this condition becomes particularly transparent when the two interacting modes belong to orthogonal polarization families, such as the TE and TM modes.

For a WGM resonator, the azimuthal dependence of an optical mode can be written approximately as
\begin{equation}
    {\bf E}_{j}({\bf r}) \propto
    {\bf e}_{j}({\bf r})\exp(i L_j\phi),
\end{equation}
where $L_j$ is the orbital mode number.  The electro-optic interaction with a microwave field having an azimuthal dependence characterized by $L_{\mathrm{RF}}$ then contains an overlap integral of the general form
\begin{equation}
    g_{\mathrm{eo}} \propto
    \int_V {\bf E}_{1}^{*}\cdot
    \left(\delta\boldsymbol{\epsilon}_{\mathrm{eo}}
    {\bf E}_{2}\right)\,dV ,
\end{equation}
where 
$\delta\boldsymbol{\epsilon}_{\mathrm{eo}}$ is the microwave-field-induced perturbation of the optical dielectric tensor.  The azimuthal part of this integral imposes a momentum, or equivalently angular momentum, conservation condition,
\begin{equation}
    |L_2-L_1|=L_{\mathrm{RF}} .
    \label{eq:eo_selection}
\end{equation}
The absolute value of the difference is needed if one considers standing microwave field. In the case of the running waves, the absolute value should be elevated. The spatial periodicity of the microwave field must be compatible with the difference between the azimuthal propagation constants of the two optical WGMs.

For a conventional travelling-wave electro-optic interaction, this condition can also be expressed in terms of the propagation constants. Let us consider the case when TM mode is optically pumped and as the result of the opto-electronic interaction TE and RF photons are generated. In this case
\begin{equation}
    k_{\mathrm{TM}}-k_{\mathrm{TE}}
    \simeq k_{\mathrm{RF}},
    \label{eq:phase_matching}
\end{equation}
where the effective optical propagation constants are $k_j=n_j\omega_j/c$.  Since the microwave wavelength is generally much longer than the optical wavelength, the required difference between the optical propagation constants is relatively small compared
with either optical propagation constant individually.  Consequently, the refractive-index difference between the two optical polarization families becomes a critical parameter.

In an isotropic dielectric, the TE and TM modes experience essentially
the same refractive index.  Their frequency splitting is then governed primarily by geometrical and polarization-dependent corrections. Although these effects can produce a finite separation between the
modes, they do not provide an independent and readily controllable mechanism for matching the optical momentum difference to the microwave wavevector.  In a birefringent material, in contrast, the two
polarizations experience different refractive indices,
\begin{equation}
    n_{\mathrm{TE}}\neq n_{\mathrm{TM}},
\end{equation}
and therefore acquire different optical propagation constants,
\begin{equation}
    k_{\mathrm{TE}} =
\frac{n_{\mathrm{TE}}\omega_{\mathrm{TE}}}{c},
    \qquad
    k_{\mathrm{TM}} =
\frac{n_{\mathrm{TM}}\omega_{\mathrm{TM}}}{c}.
\end{equation}
The resulting birefringent splitting can be used to bring the TE--TM frequency difference into coincidence with the microwave frequency. This is especially important for a resonator in which the microwave field is represented by a low-order dielectric mode or by a near-field electrode structure.

The phase matching also should satisfy the resonant frequency matching condition. By selecting the resonator radius, crystallographic orientation, and optical mode numbers, one can arrange
\begin{equation}
    \omega_{\mathrm{TM}}-\omega_{\mathrm{TE}}
    \simeq \Omega_{\mathrm{RF}},
    \label{eq:frequency_matching}
\end{equation}
while simultaneously satisfying the spatial selection rule
of Eq.~(\ref{eq:phase_matching}).

The relaxed condition for ideal phase matching should be written with respect of the resonator's radius. Instead of Eq.~(\ref{eq:phase_matching}) one can use
\begin{equation} \label{eq:soft}
    2\pi R \left | \frac{n_{TM}-n_{TE}}{\lambda_{TE}} + \frac{n_{TM}-n_{RF}}{\lambda_{RF}} \right | \ll 1.
\end{equation}
Since $\lambda_{RF}$ usually exceeds $\lambda$ by several orders of magnitude, the phase matching is achieved in the vicinity of $n_{TM} \approx n_{TE}$.

This consideration has direct implications for receiver sensitivity. The electro-optic frequency-conversion efficiency is proportional to the squared magnitude of the mode-overlap integral, and hence a phase-matched interaction can produce a much larger optical sideband
for a given incident microwave field. This observation also explains why anisotropic crystalline resonators with low birefringence like lithium tantalate are particularly attractive for microwave-to-optical conversion. The birefringence can be selected through the crystal and its
orientation, allowing the TE and TM dispersion curves to be engineered so that their frequency difference and azimuthal momentum difference simultaneously satisfy the microwave matching conditions.  In this sense, birefringence should be regarded as a resource for electro-optic receiver design rather than simply as a perturbation to the optical
spectrum.  Properly engineered birefringence can transform the
polarization degree of freedom of the WGM resonator into the mechanism that establishes phase matching between the optical and microwave fields, thereby enabling efficient frequency conversion and improving the attainable electric-field sensitivity. 

\section{Conclusions}

High-$Q$ WGM resonators are excellent tools for studying optical material properties that affect the index of refraction. We already mentioned our earlier study of near-infrared-induced photorefraction in lithium niobate resonators \cite{savchenkov06photorefraction,savchenkov06MgOLiNbO3,savchenkov07damage} conducted through this technique. In this work we demonstrated its application to accurately characterising a weak and possibly sign-changing birefringence in optical crystals. This was done with a very common and important nonlinear-optical crystal, the LTA. In spite of its prevalence, the birefringence of this material was not well characterized. There were conflicting accounts even regarding its sign. Our measurements allowed us to determine $\Delta n=(1.35\pm 0.09)\times 10^{-3}$ at room temperature and 1550 nm wavelength in a large resonator fabricated at University of Otago from stoichiometric LTA. This result is close to $\Delta n=(1.8\pm 0.1)\times 10^{-3}$ found at 1558 nm in a small resonator fabricated at JPL from the same material, and calculated using a different approach. Note that different LTA wafers have been used, and the test temperatures may differ by a few degrees. This unambiguously indicates that this crystal has negative birefringence. On the other hand, for a congruent LTA we found $\Delta n=(-5.2\pm 0.1)\times 10^{-3}$ which makes it a positively birefringent crystal. Similar measurements can be performed for various wavelengths and temperatures, leading to a thorough characterization of this crystal optical properties and an insight into the underlying solid-state physics.

Understanding and control of the weak birefringence in LTA is important for the phase matching of various nonlinear conversion processes. Specifically for the WGM resonator research, the ability to continuously tune the birefringence around zero opens up a possibility to design a resonator with equal or nearly-equal TE and TM FSRs. This would be desirable in tunable nonlinear frequency converter of Type-II, such that couple different polarization of optical signals. Tuning across positive-zero-negative birefringence would open particularly interesting perspectives in the x-cut and angle-cut geometries, when the optical axis lies in the WGM plane or makes with it a non-trivial angle \cite{Lin2012angle-cut,Sedlmeir13angle}, respectively.

\begin{backmatter}
%\bmsection{Funding}
%Content in the funding section will be generated entirely from details submitted to Prism. Authors may add placeholder text in this section to assess length, but any text added to this section will be replaced during production and will display official funder names along with any grant numbers provided. If additional details about a funder are required, they may be added to the Acknowledgment, even if this duplicates some information in the funding section. For preprint submissions, please include funder names and grant numbers in the manuscript.

\bmsection{Acknowledgment}
Research conducted by ABM and AAS was carried out at the Jet Propulsion Laboratory,
California Institute of Technology, under a contract
with the National Aeronautics and Space Administration
(80NM0018D0004).
Research conducted by FS was funded by Horizon Europe, Marie Sklodowska-Curie Actions (MSCA) (Grant No. 101154422) and Marsden Fast Start (24-UOO-196) of the Te Apārangi Royal Society of New Zealand. LS and NJL are supported by Quantum Technologies Aotearoa, a research program funded by the New Zealand Ministry of Business, Innovation and Employment, contract no.~UOO2347. The authors thank Vladimir S. Ilchenko for discussions.

\bmsection{Disclosures}
The authors declare no conflicts of interest.

\bmsection{Data Availability Statement}
No data were generated or analyzed in the presented research.

\end{backmatter}

%%%%%%%%%%%%%%%%%%%%%%% References %%%%%%%%%%%%%%%%%%%%%%%%%

%\bibliography{bibliography}

\begin{thebibliography}{10}
\newcommand{\enquote}[1]{``#1''}

\bibitem{Abedin1996LiTaO3}
K.~S. Abedin and H.~Ito, \enquote{Temperature-dependent dispersion relation of ferroelectric lithium tantalate,} {\protect\JournalTitle{Journal of Applied Physics}} \textbf{80}, 6561--6563 (1996).

\bibitem{nakamura2002SLT}
M.~Nakamura, S.~Higuchi, S.~Takekawa, \emph{et~al.}, \enquote{Refractive {{Indices}} in {{Undoped}} and {{MgO-Doped Near-Stoichiometric LiTaO3 Crystals}},} {\protect\JournalTitle{Japanese Journal of Applied Physics}} \textbf{41}, L465--L467 (2002).

\bibitem{kostritskii2009LTA}
S.~M. Kostritskii, M.~Aillerie, P.~Bourson, and D.~Kip, \enquote{Raman spectroscopy study of compositional inhomogeneity in lithium tantalate crystals,} {\protect\JournalTitle{Appl. Phys. B}} \textbf{95}, 125--130 (2009).

\bibitem{Dolev2009LTa}
I.~Dolev, A.~Ganany-Padowicz, O.~Gayer, \emph{et~al.}, \enquote{Linear and nonlinear optical properties of {MgO}:{LiTaO}3,} {\protect\JournalTitle{Appl. Phys. B}} \textbf{96}, 423--432 (2009).

\bibitem{huband2017LiNbTaO3}
S.~Huband, D.~S. Keeble, N.~Zhang, \emph{et~al.}, \enquote{Crystallographic and optical study of {{LiNb}}{\textsubscript{ 1 - {\mkbibemph{x}} }} {{Ta}} {\mkbibemph{{\textsubscript{x}} }} {{O}}{\textsubscript{3}},} {\protect\JournalTitle{Acta Crystallographica Section B Structural Science, Crystal Engineering and Materials}} \textbf{73}, 498--506 (2017).

\bibitem{huband2017LTA}
S.~Huband, D.~S. Keeble, N.~Zhang, \emph{et~al.}, \enquote{Relationship between the structure and optical properties of lithium tantalate at the zero-birefringence point,} {\protect\JournalTitle{J. Appl. Phys.}} \textbf{121}, 024102 (2017).

\bibitem{strekalov2022Receiver}
D.~Strekalov, N.~Majurec, A.~Matsko, \emph{et~al.}, \enquote{W-{{Band Photonic Receiver}} for {{Compact Cloud Radars}},} {\protect\JournalTitle{Sensors}} \textbf{22}, 804 (2022).

\bibitem{savchenkov06photorefraction}
A.~A. Savchenkov, A.~B. Matsko, D.~Strekalov, \emph{et~al.}, \enquote{Enhancement of photorefraction in whispering gallery mode resonators,} {\protect\JournalTitle{Phys. Rev. B}} \textbf{74}, 245119 (2006).

\bibitem{savchenkov06MgOLiNbO3}
A.~A. Savchenkov, A.~B. Matsko, D.~Strekalov, \emph{et~al.}, \enquote{Photorefractive effects in magnesium doped lithium niobate whispering gallery mode resonators,} {\protect\JournalTitle{Appl. Phys. Lett.}} \textbf{88}, 241909 (2006).

\bibitem{savchenkov07damage}
A.~A. Savchenkov, A.~B. Matsko, D.~Strekalov, \emph{et~al.}, \enquote{Photorefractive damage in whispering gallery resonators,} {\protect\JournalTitle{Opt. Comm.}} \textbf{272}, 257--262 (2007).

\bibitem{Li12sb}
J.~Li, H.~Lee, K.~Y. Yang, and K.~Vahala, \enquote{Sideband spectroscopy and dispersion measurement in microcavities,} {\protect\JournalTitle{Opt. Expr.}} \textbf{20}, 26337--26344 (2012).

\bibitem{Savchenkov12overmoded}
A.~A. Savchenkov, A.~B. Matsko, W.~Liang, \emph{et~al.}, \enquote{Kerr frequency comb generation in overmoded resonators,} {\protect\JournalTitle{Opt. Expr.}} \textbf{20}, 27290--27298 (2012).

\bibitem{Strekalov14SFG}
D.~V. Strekalov, A.~S. Kowligy, Y.-P. Huang, and P.~Kumar, \enquote{Optical sum-frequency generation in a whispering-gallery-mode resonator,} {\protect\JournalTitle{New J. Phys.}} \textbf{16}, 053025 (2014).

\bibitem{grudinin06highQlin}
I.~S. Grudinin, V.~S. Ilchenko, and L.~Maleki, \enquote{Ultrahigh optical q factors of crystalline resonators in the linear regime,} {\protect\JournalTitle{Phys. Rev. A}} \textbf{74}, 63806 (2006).

\bibitem{grudinin06highQ}
I.~S. Grudinin, A.~B. Matsko, A.~A. Savchenkov, \emph{et~al.}, \enquote{Ultra high q crystalline microcavities,} {\protect\JournalTitle{Opt. Comm.}} \textbf{265}, 33--38 (2006).

\bibitem{fujii2020fab}
S.~Fujii, Y.~Hayama, K.~Imamura, \emph{et~al.}, \enquote{All-precision-machining fabrication of ultrahigh-{{Q}} crystalline optical microresonators,} {\protect\JournalTitle{Optica}} \textbf{7}, 694 (2020).

\bibitem{minkov2021Fabrication}
K.~N. Min'kov, G.~V. Likhachev, N.~G. Pavlov, \emph{et~al.}, \enquote{Fabrication of high-{{Q}} crystalline whispering gallery mode microcavities using single-point diamond turning,} {\protect\JournalTitle{J. Opt. Technol.}} \textbf{88}, 348 (2021).

\bibitem{Demchenko13}
Y.~A. Demchenko and M.~L. Gorodetsky, \enquote{Analytical estimates of eigenfrequencies, dispersion, and field distribution in whispering gallery resonators,} {\protect\JournalTitle{JOSA B}} \textbf{30}, 3056--3063 (2013).

\bibitem{Gorodetsky99coupling}
M.~L. Gorodetsky and V.~S. Ilchenko, \enquote{Optical microsphere resonators: optimal coupling to high-q whispering-gallery modes,} {\protect\JournalTitle{J. Opt. Soc. Am. B}} \textbf{16}, 147--154 (1999).

\bibitem{foreman2016dielectric}
M.~R. Foreman, F.~Sedlmeir, H.~G.~L. Schwefel, and G.~Leuchs, \enquote{Dielectric tuning and coupling of whispering gallery modes using an anisotropic prism,} {\protect\JournalTitle{Journal of the Optical Society of America B}} \textbf{33}, 2177 (2016).

\bibitem{sedlmeir_polarization-selective_2017}
F.~Sedlmeir, M.~R. Foreman, U.~Vogl, \emph{et~al.}, \enquote{Polarization-{{Selective Out}}-{{Coupling}} of {{Whispering}}-{{Gallery Modes}},} {\protect\JournalTitle{Phys. Rev. Appl.}} \textbf{7}, 024029 (2017).

\bibitem{trainor2018selective}
L.~S. Trainor, F.~Sedlmeir, C.~Peuntinger, and H.~G.~L. Schwefel, \enquote{Selective {{Coupling Enhances Harmonic Generation}} of {{Whispering-Gallery Modes}},} {\protect\JournalTitle{Physical Review Applied}} \textbf{9}, 024007 (2018).

\bibitem{Schunk14modes}
G.~Schunk, J.~U. F{\"u}rst, M.~F{\"o}rtsch, \emph{et~al.}, \enquote{Identifying modes of large whispering-gallery mode resonators from the spectrum and emission pattern.} {\protect\JournalTitle{Opt. Expr.}} \textbf{22}, 30795--30806 (2014).

\bibitem{Oraevsky02WGM}
A.~N. Oraevsky, \enquote{Whispering-gallery waves,} {\protect\JournalTitle{Quant. El.}} \textbf{35}, 377--400 (2002).

\bibitem{Breunig13eigenfunc}
I.~Breunig, B.~Sturman, F.~Sedlmeir, \emph{et~al.}, \enquote{Whispering gallery modes at the rim of an axisymmetric optical resonator: analytical versus numerical description and comparison with experiment.} {\protect\JournalTitle{Opt. Expr.}} \textbf{21}, 30683--30692 (2013).

\bibitem{Lin2012angle-cut}
G.~Lin, J.~F\"urst, D.~V. Strekalov, \emph{et~al.}, \enquote{High-{{Q UV}} whispering gallery mode resonators made of angle-cut {{BBO}} crystals,} {\protect\JournalTitle{Optics Express}} \textbf{20}, 21372 (2012).

\bibitem{Sedlmeir13angle}
F.~Sedlmeir, M.~Hauer, J.~U. F{\"u}rst, \emph{et~al.}, \enquote{Experimental characterization of an uniaxial angle cut whispering gallery mode resonator.} {\protect\JournalTitle{Opt. Expr.}} \textbf{21}, 23942--23949 (2013).

\end{thebibliography}

\end{document}